\documentclass[a4paper,11pt]{article}

\usepackage{pos}

\newcommand{\Mpl}{M_{\rm pl}}
\title{Recasting scalar-tensor theories of gravity for colliders}

\usepackage{braket}

\author[a]{Andrei Lazanu}
\author*[a]{Peter Millington}
\author[b]{Sergio Sevillano Muñoz}

\affiliation[b]{Department of Physics and Astronomy, University of Manchester, Oxford Road, Manchester M13 9PL, United Kingdom}

\affiliation[b]{Institute for Particle Physics Phenomenology, Department of Physics, Durham University, Durham DH1 3LE, United Kingdom}

\emailAdd{andrei.lazanu@manchester.ac.uk}
\emailAdd{peter.millington@manchester.ac.uk}
\emailAdd{sergio.sevillano-munoz@durham.ac.uk}

\abstract{Diagrammatic approaches to perturbation theory transformed the practicability of calculations in particle physics. In the case of extended theories of gravity, however, obtaining the relevant diagrammatic rules is non-trivial:\ we must expand in metric perturbations and around (local) minima of the scalar field potentials, make multiple field redefinitions, and diagonalize kinetic and mass mixings. In this note, we will motivate these theories, introduce the package \texttt{FeynMG} --- a \texttt{Mathematica} extension of \texttt{FeynRules} that automates the process described above --- and highlight an application to a model with unique collider phenomenology.\\
\smallskip

IPPP/24/70}

\FullConference{42nd International Conference on High Energy Physics (ICHEP2024)\\
18-24 July 2024\\
Prague, Czech Republic\\}

\begin{document}
\maketitle

%%%%%%%%%%%%%%%%%%%%

\section{Introduction}

Dark matter (DM) remains a strong motivation for ultra-light modifications to the Standard Model (SM). Converting the virial radius of a galaxy (say, around $10$ kPc) to a mass scale suggests a lower limit on the DM mass of order $10^{-22}$ eV (see Ref.~\cite{Zimmermann:2024xvd} for more robust bounds). Similarly light scalar fields can lead to new long-range forces or \textit{fifth forces}, and if we want to modify dynamics on even larger scales, we might consider even smaller masses. These new scalar fields can be coupled directly to the SM or introduced by modifying the Einstein--Hilbert action of general relativity (GR)
\begin{equation}
    S=\int\!{\rm d}^4x\;\sqrt{-g}\,\frac{M_{\rm Pl}^2}{2}\left(R-2\Lambda\right)\;,
\end{equation}
where $M_{\rm Pl}$ is the (reduced) Planck mass, $R$ is the Ricci scalar and $\Lambda$ is the cosmological constant.

In this note, we start with the simplest scalar-tensor theory, where the Planck mass is promoted to a dynamical scalar field. We describe how fifth-force couplings arise and illustrate the role played by scale symmetry breaking. Next, we introduce the package \texttt{FeynMG}~\cite{SevillanoMunoz:2022tfb}, which can automate the calculation of the couplings between the fifth-force scalar and SM, and highlight its use in the analysis of a model in which the scalar couples only to off-shell SM degrees of freedom.

%%%%%%%%%%%%%%%%%%%%

\section{Fifth forces and screening}
\label{sec:fifthforces}

We can redefine the rulers that we use to measure distances in spacetime by rescaling the metric $g_{\mu\nu}$. This can be done locally with a Weyl rescaling of the form $g_{\mu\nu}\to \Omega^2(x)g_{\mu\nu}$. Under this rescaling, and with a simultaneous redefinition of the affine parameter $\lambda$ via \smash{$\frac{{\rm d}}{{\rm d}\lambda}\to \Omega^{-2}(x)\,\frac{\rm d}{{\rm d}\lambda}$}, the geodesic equation transforms as
\begin{equation}
    \label{eq:geodesic}
    \frac{{\rm d}^2x^{\mu}}{{\rm d}\lambda^2}+\Gamma^{\mu}_{\alpha\beta}\,\frac{{\rm d}x^{\alpha}}{{\rm d}\lambda}\,\frac{{\rm d}x^{\beta}}{{\rm d}\lambda}=0\quad \longrightarrow\quad \frac{{\rm d}^2x^{\mu}}{{\rm d}\lambda^2}+\Gamma^{\mu}_{\alpha\beta}\,\frac{{\rm d}x^{\alpha}}{{\rm d}\lambda}\,\frac{{\rm d}x^{\beta}}{{\rm d}\lambda}-g_{\alpha\beta}\,\frac{{\rm d}x^{\alpha}}{{\rm d}\lambda}\,\frac{{\rm d}x^{\beta}}{{\rm d}\lambda}\,\frac{\partial}{\partial x_{\mu}}\ln\Omega=0\;,
\end{equation}
where $\Gamma^{\mu}_{\alpha\beta}$ are the Christoffel symbols. The Weyl rescaling has led to an extra force term proportional to the gradient of $\Omega$. However, if the geodesic is null, i.e., \smash{$g_{\alpha\beta}\,\frac{{\rm d}x^{\alpha}}{{\rm d}\lambda}\,\frac{{\rm d}x^{\beta}}{{\rm d}\lambda}=0$}, this fifth-force term vanishes, hinting at the dependence of fifth-force couplings on how scale symmetry is broken.

Weyl rescaling allows us to shuffle scalar couplings in our scalar-tensor theory of gravity between the gravitational and non-gravitational sectors. Starting with a \textit{Jordan-frame} action
\begin{equation}\label{eq:Jordan action}
    S=\int\!{\rm d}^4x\;\sqrt{-g}\,\bigg[\frac{F(\phi)}{2}R-\frac{Z^{\mu\nu}(\phi,\partial \phi,\dots)}{2}\partial_{\mu}\phi\partial_{\nu}\phi-V(\phi)+\mathcal{L}_{\rm SM}(g_{\alpha\beta},\{\psi\})\bigg]\;,
\end{equation}
we can move to a frame in which the gravitational action is of Einstein--Hilbert form by the Weyl rescaling
$g_{\mu\nu}=M_{\rm Pl}^2F^{-1}(\phi)\tilde{g}_{\mu\nu}=A^2(\tilde{\phi})\tilde{g}_{\mu\nu}$, leading to the \textit{Einstein-frame} action
\begin{equation}
    S=\int\!{\rm d}^4x\;\sqrt{-\tilde{g}}\,\bigg[\frac{M_{\rm Pl}^2}{2}\tilde{R}-\frac{\tilde{Z}^{\mu\nu}(\tilde{\phi},\partial \tilde{\phi},\dots)}{2}\partial_{\mu}\tilde{\phi}\partial_{\nu}\tilde{\phi}-\tilde{V}(\tilde{\phi})+\mathcal{L}_{\rm SM}(A^2(\tilde{\phi})\tilde{g}_{\alpha\beta},\{\psi\})\bigg]\;,
\end{equation}
wherein we see that the SM degrees of freedom move on geodesics that are modified with respect to the Einstein-frame metric $\tilde{g}_{\mu\nu}$ by the coupling function $A^2(\tilde{\phi})$. Note that we use the $(-,+,+,+)$ signature convention. When the coupling function is close to unity, we can make a Taylor expansion
\begin{equation}
    S_{\rm SM}[A^2(\tilde{\phi})\tilde{g}_{\mu\nu},\{\psi\}]=S_{\rm SM}[\tilde{g}_{\mu\nu},\{\psi\}]+\frac{1}{2}[A^2(\tilde{\phi})-1]\tilde{T}_{\rm SM}+\dots\;,
\end{equation}
and we obtain a ``universal'' coupling of the scalar to the trace of the SM energy-momentum tensor $\tilde{T}_{\rm SM}$. As a result, test particles will experience a force per unit mass $\vec{F}/m=-\vec{\nabla}\ln A(\tilde{\phi})$, cf.~Eq.~\eqref{eq:geodesic}.

Long-range fifth forces are heavily constrained by local tests of gravity. To find ways of evading these constraints, we can consider the fifth-force potential of a point mass $\mathcal{M}$ due to the exchange of fluctuations in the scalar field around some background $\tilde{\varphi}=\braket{\tilde{\phi}}$. The potential takes the form~\cite{Joyce:2014kja}
\begin{equation}
    \tilde{U}(r)= -\frac{1}{\tilde{Z}(\tilde{\varphi})c_s^2(\tilde{\varphi})}\left[\frac{{\rm d}A(\tilde{\varphi})}{{\rm d}\tilde{\varphi}}\right]^2\frac{1}{4\pi r}\exp\left[-\,\frac{m(\tilde{\varphi})r}{\tilde{Z}^{1/2}(\tilde{\varphi})c_s(\tilde{\varphi})}\right]\mathcal{M}\;.
\end{equation}
The force can be suppressed by (see Ref.~\cite{Joyce:2014kja}):\ (i) increasing the mass $m(\tilde{\varphi})$, (ii) increasing the sound speed $c_s(\tilde{\varphi})$ or the pre-factor of the kinetic term $\tilde{Z}(\tilde{\varphi})$ (both encoded in $\tilde{Z}^{\mu\nu}$ above) --- or decreasing the same to enhance the Yukawa suppression --- or (iii) decreasing the coupling strength ${\rm d}A(\tilde{\varphi})/{\rm d}\tilde{\varphi}$. These factors depend on the background field $\tilde{\varphi}$, and thereby the ambient matter density, leading to environmental screening of the force, and common types are chameleon~\cite{Khoury:2003aq} [via (i) above], Vainshtein~\cite{Vainshtein:1972sx} [increasing $Z(\tilde{\varphi})$ via (ii) above] and symmetron~\cite{Gessner:1992flm, Damour:1994zq, Hinterbichler:2010es} [via (iii) above] screening.

\begin{figure}[!t]
    \centering
    \includegraphics[width=0.95 \textwidth]{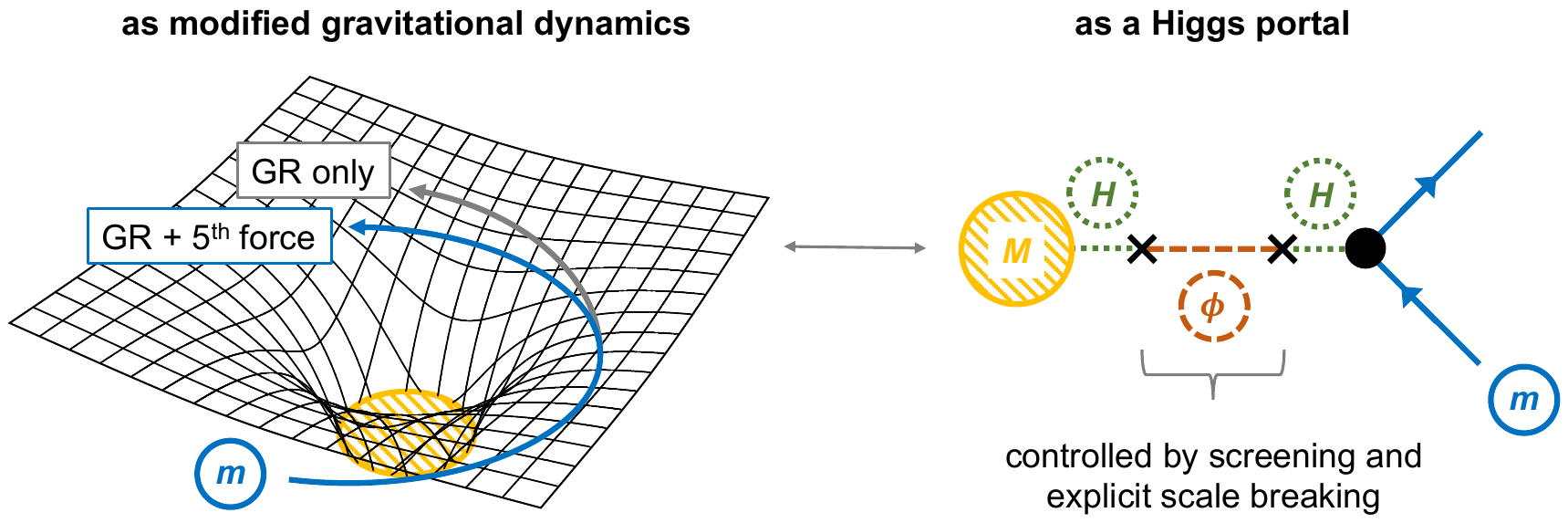}
    \caption{Equivalent descriptions of fifth forces exerted by a mass $M$ on a test mass $m$ in the SM, as modified gravitational dynamics (left) or as Higgs-portal-induced mixing with the Higgs boson (right).\vspace{-0.5em}}
    \label{fig:fifthforce}
\end{figure}

However, from Eq.~\eqref{eq:geodesic}, we expect that how fifth forces couple to the SM also depends on how scale symmetry is broken. In the SM (excluding Majorana mass terms), there is one explicit scale-breaking term:\ the quadratic term in the Higgs potential $
\mathcal{L}_{\rm SM}\supset -\mu^2|H|^2$. It is only this term to which the scalar will couple (classically) after the Weyl rescaling, meaning that a coupling $\phi^2 R$ is equivalent to a Higgs-portal coupling $\phi^2|H|^2$ (up to dimension-four operators)~\cite{Burrage:2018dvt} (see Fig.~\ref{fig:fifthforce}). The coupling to SM leptons, e.g., then arises either by resumming the mixing between the SM Higgs field and the fifth-force scalar or by diagonalizing this mixing in terms of the light eigenmode $\zeta$, giving a coupling \smash{$-\frac{2\mu^2}{m_H^2}\frac{m_L}{M}\bar{\psi}_L\zeta\psi_L$}~\cite{Burrage:2018dvt}. Here, $m_L$ and $m_H$ are the fermion and Higgs-boson masses, respectively, and $M$ is the mass scale that determines the strength of the fifth-force coupling. For the SM, the ratio $2\mu^2/m_H^2=1$, and we recover the usual ``universal'' coupling, described earlier~\cite{Burrage:2018dvt}. In cases where $\mu\to 0$ and the Higgs-boson mass is generated through dynamical scale symmetry breaking, this ratio and the fifth-force coupling vanish~\cite{Shaposhnikov:2008xb, Brax:2014baa, Ferreira:2016kxi}.

The coupling to the elementary fermions was mediated by an explicit scale-breaking term, but the SM has additional dynamical sources of scale breaking. In hadrons, gluons contribute to the trace of the energy-momentum tensor through an anomaly, and the fifth-force coupling to nucleons, e.g., is $-\eta\frac{m_N}{M}\bar{\psi}_N\zeta\psi_N$~\cite{Burrage:2018dvt}. It can be shown that the parameter $\eta<1$, suggesting that there may be effective violations of the universality of free fall between elementary and composite states~\cite{Burrage:2018dvt}.

%%%%%%%%%%%%%%%%%%%%

\section{\texttt{FeynMG} and the Jordan frame}

\begin{figure}
    \centering
    \includegraphics[width=0.9\linewidth]{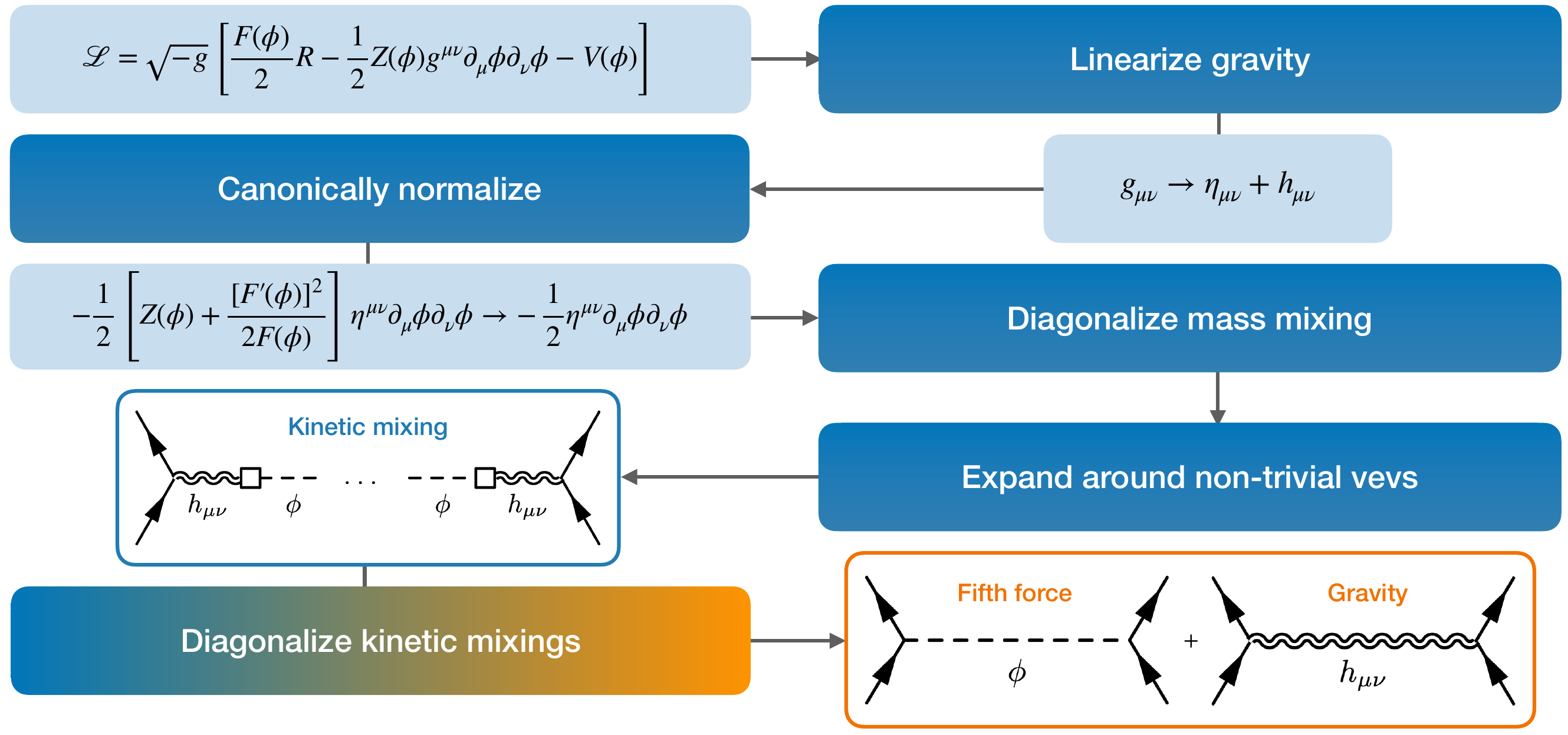}
    \caption{Schematics of the procedure to determine scalar-mediated interactions from the Jordan-frame description of a non-minimally coupled scalar field [see Eq.~\eqref{eq:Jordan action}], as automated by \texttt{FeynMG}~\cite{SevillanoMunoz:2022tfb}.\vspace{-0.5em}}
    \label{fig: FeynMG flowchart}
\end{figure}

Section~\ref{sec:fifthforces} was developed in the Einstein frame. However, there are theories for which an Einstein frame does not exist. We must then work in the Jordan-frame with a non-trivial gravitational sector, and the procedure for determining the scalar-mediated interactions (see Ref.~\cite{Copeland:2021qby}) is shown in Fig.~\ref{fig: FeynMG flowchart}.

The first step is to linearize gravity around a Minkowski background. To this end, we expand \smash{$g_{\mu\nu}\to \eta_{\mu\nu}+ h_{\mu\nu}$} and \smash{$g^{\mu\nu}\to \eta^{\mu\nu}- h^{\mu\nu}+\dots $}, where $\eta_{\mu\nu}$ is the Minkowski metric and $h_{\mu\nu}$ is the graviton field. The omitted terms ensure that \smash{$g_{\mu\rho}g^{\rho\nu}=\delta_\mu^{\phantom{\mu}\nu}$}. Applying this linearization to Eq.~\eqref{eq:Jordan action}, we obtain
\begin{align}
    \mathcal{L} & = \frac{F(\phi)}{4}\bigg[\frac{1}{4}\partial_{\mu}h\partial^{\mu}h-\frac{1}{2}\partial_{\rho}h_{\mu\nu}\partial^{\rho}h^{\mu\nu}\bigg]+\frac{F'(\phi)}{4}\eta^{\mu\nu}\partial_{\mu}h\partial_{\nu}\phi-\frac{1}{2}\bigg[Z(\phi)+\frac{[F'(\phi)]^2}{2F(\phi)}\bigg]\eta^{\mu\nu}\partial_{\mu}\phi\partial_{\nu}\phi\nonumber\\
    &\phantom{=}-V(\phi)+\frac{1}{2}h_{\mu\nu}T^{\mu\nu}_{\rm SM}+\mathcal{L}_{\rm SM}(\eta,\{\psi\})+\dots\;,\quad \text{with}\quad h\equiv \eta^{\mu\nu}h_{\mu\nu}\;,
\end{align}
to leading order in \smash{$F^{-1/2}(\phi)$}. Note that the Planck mass is determined by the vacuum expectation value (vev) $v_{\phi}=\braket{\phi}$ via $\Mpl^2=F(v_\phi)$. It is the resulting kinetic mixing between the graviton and scalar (second term) that can lead to fifth forces. To see this, we redefine the scalar and graviton fields following the steps in Fig.~\ref{fig: FeynMG flowchart}, diagonalizing the kinetic mixing and leading to the Lagrangian
\begin{align}
    \label{eq: Jordan canon}
    \mathcal{L} &=\frac{1}{4} \partial_\mu h\partial^\mu h-\frac{1}{2}\partial_\rho h_{\mu\nu}\partial^\rho h^{\mu\nu}
	-\frac{1}{2}\partial_\mu \phi \partial^\mu \phi -{V}(v_{\phi}+\phi)\nonumber\\
	&\phantom{=}+h_{\mu\nu}T_{\rm SM}^{\mu\nu}+\frac{1}{2}\hat{F}'(v_\phi)\big(1+\hat{F}'(v_\phi)^2\big)^{-1/2}\phi\, T_{\rm SM}+\mathcal{L}_{\rm SM}(\eta,\{\psi\})+\dots\;,\quad \text{with}\quad M_{\rm Pl}= 1\;.
\end{align}

This process can be automated using the \texttt{Mathematica} package \texttt{FeynMG}~\cite{SevillanoMunoz:2022tfb}, which acts as an extension to \texttt{FeynRules}~\cite{Alloul:2013bka}. \texttt{FeynMG} allows the user to take any matter Lagrangian defined in Minkowski spacetime, automatically incorporate all minimal couplings to gravity, and append any scalar-tensor gravitational sector. From there, \texttt{FeynMG} contains the necessary functions to express the new interactions arising from the modification of gravity in terms of an effective extension of the SM, which can be analyzed further using standard packages.

\begin{figure}
    \centering
    \includegraphics[width=0.48\textwidth]{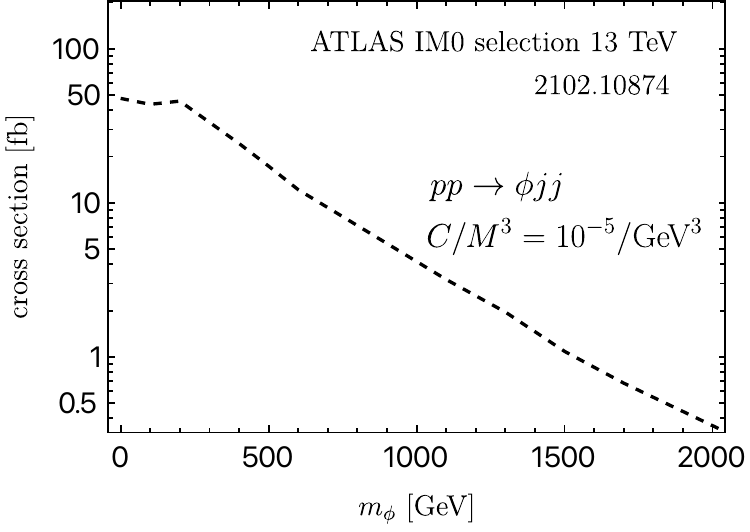} \includegraphics[width=0.46\textwidth]{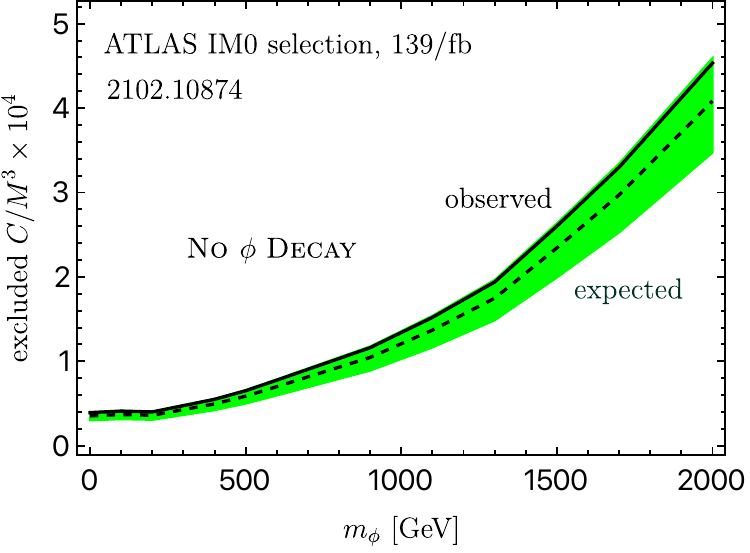}
    \caption{Cross section for $pp$ to $\phi$ plus two jets as a function of the scalar mass $m_{\phi}$ and the constraints on the coupling $C/M^3$ (right), assuming the scalar to be be stable, reproduced from Ref.~\cite{Englert:2024ryd}.\vspace{-0.5em}}
    \label{fig:cross-section}
\end{figure}

For example, \texttt{FeynMG} can be used to obtain the energy-momentum tensor from a given \texttt{FeynRules} model file and introduce a coupling of the form~\cite{Englert:2024ryd}
\begin{equation}
    \label{eq:offshell_coupling}
    \mathcal{L}\supset -(C/M^3)T^{\mu\nu}_{\rm SM}\partial_{\mu}\partial_{\nu}\phi\;.
\end{equation}
Since the energy-momentum tensor of an on-shell SM state has vanishing divergence, the scalar will couple only to off-shell states. This means that the first deviations will arise in $2\to 3$ or loop-level processes. The coupling in Eq.~\eqref{eq:offshell_coupling} was analyzed in Ref.~\cite{Englert:2024ryd} in the context of an LHC mono-jet analysis. Using the inclusive ‘IM0’ search region of Ref.~\cite{ATLAS:2021kxv} as a proxy, Ref.~\cite{Englert:2024ryd} obtained cross sections (left) and constraints on $C/M^3$ (under the assumption that the scalar is stable) as a function of the $\phi$ mass (right), shown in Fig.~\ref{fig:cross-section}.

%%%%%%%%%%%%%%%%%%%%

\section{Closing remarks}

We have given a  brief overview of the screened fifth-force couplings that can arise in scalar-tensor theories of gravity. We have highlighted the role of scale symmetry breaking in determining the form of these couplings. As a result, we have seen that there are parallels between scalar-tensor theories of gravity and Higgs-portal theories, and therefore, more generally, other models involving ultra-light scalar fields. Last, we have shown how the analysis of scalar-mediated processes arising in scalar-tensor theories can be automated with \texttt{FeynRules} by using the package \texttt{FeynMG}. We have highlighted an example of its use for a model in which the scalar couples only to off-shell SM states, leading to a unique collider phenomenology that impacts only higher-multiplicity processes.

%%%%%%%%%%%%%%%%%%%%

\section*{Acknowledgements}

The authors thank Clare Burrage, Ed Copeland, Christoph Englert and Michael Spannowsky for their collaboration. This work was supported by the Science and Technology Facilities Council (STFC) via [Grant No.~ST/X00077X/1] and [Grant No.~ST/T001011/1], and a United Kingdom Research and Innovation (UKRI) Future Leaders Fellowship [Grant No.~MR/V021974/2].

\paragraph{Data Access Statement} Data and codes supporting the works summarized in this note can be accessed via the references cited in the text.

%%%%%%%%%%%%%%%%%%%%

\end{document}